\documentclass[twocolumn]{aastex63}
\usepackage{natbib}
\usepackage{multirow}
\usepackage{gensymb}
\usepackage{lipsum}
\usepackage{blindtext}
\usepackage{mathtools}
\usepackage{tensor}
\emergencystretch=\maxdimen
\shorttitle{Investigating LQG Motivated Rotating Black Holes with EHT Observation}
\shortauthors{S.~U.~Islam et al.}
\definecolor{MyDarkBlue}{rgb}{0,0.08,0.5}
\definecolor{MyDarkRed}{rgb}{0.7,0.02,0.02}
\definecolor{MyDarkGreen}{rgb}{0.0,0.7,0.0}

\usepackage{amsmath}

\begin{document}
	%------------------------------------------------------------------------------
	% Title
	%------------------------------------------------------------------------------
\title{Investigating Loop Quantum Gravity with EHT Observational  Effects of Rotating Black holes}
	%
	%------------------------------------------------------------------------------
	% Authors
	%------------------------------------------------------------------------------
	
	\correspondingauthor{Shafqat Ul Islam}
	\email{Shafphy@gmail.com}
\author[0000-0002-8539-5755]{Shafqat Ul Islam}
\affiliation{ Centre for Theoretical Physics, Jamia Millia Islamia, New Delhi 110025, India}
\author[0000-0002-2724-9733]{Jitendra Kumar}
\affiliation{ Centre for Theoretical Physics, Jamia Millia Islamia, New Delhi 110025, India}
\author[0000-0003-2471-1360]{Rahul Kumar Walia}
\affiliation{ Astrophysics Research Centre, School of Mathematics, Statistics and Computer Science, University of KwaZulu-Natal, Private Bag 54001, Durban 4000, South Africa}
\author[0000-0002-0835-3690]{Sushant G. Ghosh}
\affiliation{ Centre for Theoretical Physics, Jamia Millia Islamia, New Delhi 110025, India}
\affiliation{ Astrophysics Research Centre, School of Mathematics, Statistics and Computer Science, University of KwaZulu-Natal, Private Bag 54001, Durban 4000, South Africa}

%------------------------------------------------------------------------------
% Abstract
%------------------------------------------------------------------------------ 
\begin{abstract}
A mathematically consistent rotating black hole model in loop quantum gravity (LQG) is yet lacking.
The scarcity of rotating black hole solutions in LQG substantially hampers the development of testing LQG from observations, e.g., from the Event Horizon Telescope (EHT) observations. The EHT observation revealed event horizon-scale images of the supermassive black holes Sgr A* and M87*. The EHT results are consistent with the shadow of a Kerr black hole of general relativity.   We present LQG-motivated rotating  black hole (LMRBH) spacetimes, which are regular everywhere and asymptotically encompass the Kerr black hole as a particular case. The LMRBH metric describes a multi-horizon black hole in the sense that it can admit up to three horizons, such that an extremal LMRBH, unlike the Kerr black hole, refers to a black hole with angular momentum $a>M$.
The metric, depending on the parameters,  describes (1) black holes with only one horizon (BH-I), (2) black holes with an event horizon and a Cauchy horizons (BH-II), (3)  black holes with three horizons (BH-III)  or (4) no-horizon (NH) spacetime,  which, we show, is almost ruled out by the EHT observations. We constrain the LQG parameter with the aid of the EHT shadow observational results of M87* and Sgr A*,respectively, for an inclination angle of $17\degree$ and $50\degree$.  In particular, the VLTI bound for the Sgr A*, $\delta\in (-0.17,0.01)$, constrains the parameters ($a,l$) such that for $0< l\leq 0.347851M\; (l\leq 2\times 10^6$ km), the allowed range of $a$ is $(0,1.0307M)$. Together with the EHT bounds of Sgr A$^*$ and M87$^*$ observables, our analysis concludes that a substantial part of BH-I and BH-II parameter space agrees with the EHT results of M87* and Sgr A*. While the EHT M87* results totally rule out the BH-III, but not that by Sgr A*.  
\end{abstract}


%%%%%%%%%%%%%%%%%%%%%%%%%%%%%%%%%%%%%%%%%%%%%%%%%%

%%%%%%%%%%%%%%%%% BODY OF PAPER %%%%%%%%%%%%%%%%%%


\section{Introduction}\label{Sec-1}
That the gravitational collapse of a massive star ($\geq 3.5 M$) leads to a spacetime singularity in general relativity (GR) is confirmed by the elegant theorem by Hawking and Penrose \citep{Hawking:1970zqf,Hawking:1973}). However, it is widely believed that these singularities result from a classical treatment of spacetime. By its very definition, the existence of a singularity means spacetime fails to exist, signaling a breakdown in the laws of physics. Thus, singularities must be substituted by some other objects in a more unified theory for these laws to exist and will not be present when quantum effects are considered \citep{Wheeler:1964}. While we do not yet have any complete quantum gravity theory, permitting us to explore the interior of the black hole and settle it separately, we must turn our attention to regular models motivated by quantum arguments. The first regular black hole solution was proposed by Bardeen (\citeyear{Bardeen:1968}). Bardeen  asserts that although there are horizons, there is no curvature singularity. Instead, the black hole center develops a de Sitter-like region, ultimately known as a black hole with a regular center. Thus, its maximal extension is that of the Reissner–Nordström spacetime but with a regular center \citep{Barrabes:1995nk,Bronnikov:2003gx}. Thereafter, several regular black hole models have been proposed based on Bardeen's idea, which mimics the demeanor of the  Schwarzschild black hole at large distances \citep{Poisson:1988,Dymnikova:1992ux,Barrabes:1995nk,Bronnikov:2003gx,Hayward:2005gi,Bronnikov:2005gm,Simpson:2018tsi}. Also, there have been  significant advances in the analysis and application of regular black holes \citep{Ayon-Beato:1998hmi,Bronnikov:2000vy,Hayward:2005gi,Zaslavskii:2009kp,Lemos:2011dq}. There is evidence that  loop quantum gravity (LQG) may be competent to fix the inevitable singularities in classical GR \citep{Ashtekar:2006wn,Ashtekar:2006es,Vandersloot:2006ws}. Because of the
inherent problems in solving the complete system, the emphasis has been on spherically symmetric
black holes \citep{Ashtekar:2005qt,Modesto:2005zm,Boehmer:2007ket,Campiglia:2007pb,Gambini:2008dy}.

The phase space quantization or semiclassical polymerization that maintains aspects of the discreteness of underlying spacetime suggested by  LQG turns out to be a fruitful technique to resolve the singularity issue, and has been used recently to significant effect \citep{Boehmer:2007ket,Campiglia:2007pb,Gambini:2008dy}. Different polymerizations can give qualitatively other regularized spacetimes, so it is of great interest to examine a more comprehensive class of models and methods. Peltola \& Kunstatter (\citeyear{Peltola:2009jm}), motivated by earlier works \citep{Ashtekar:2005qt,Modesto:2005zm,Boehmer:2007ket,Boehmer:2008fz,Campiglia:2007pb,Gambini:2008dy,Modesto:2008im,Modesto:2006mx}, used the effective field theory and the partially polymerized theory arguments to determine a static and spherically symmetric regular black hole \citep{Peltola:2008pa,Peltola:2009jm} that is asymptotically flat. One of the most striking features of this quantum-corrected black hole, unlike other regular black holes that have two horizons, is that it has a single horizon and it also encompasses a Schwarzschild black hole. 

The no-hair theorem embodies the unique qualities of the GR black hole, stating that the Kerr black hole \citep{Kerr:1963} is the only stationary, axially symmetric, and asymptotically flat vacuum solution to the Einstein field equations \citep{Israel:1967,Israel:1967za,Carter:1971zc,Hawking:1971vc,Robinson:1975bv}. The no-hair theorem suggests that astrophysical black hole candidates are Kerr black holes, but it still lacks definitive proof, and its exact nature has not yet been verified. It opens an arena for investigating properties of black holes that differ from Kerr's black hole.

Also, the spherical black hole cannot be sampled by astrophysical observations as rotating black holes are typically found in nature. The black hole spin plays a fundamental role in any astrophysical process. Further, the lack of rotating black hole models in LQG hinders the progress of testing LQG theory from observations. This has encouraged us to consider rotating or axisymmetric generalization of the spherical metric \citep{Peltola:2009jm} recently obtained by Walia (\citeyear{KumarWalia:2022ddq}). It is a Kerr-like metric, derived through the revised Newman-Janis algorithm (NJA), and hereafter called as the LQG-motivated rotating black hole (LMRBH), which is suitable to test with astrophysical observations. Also, it turns out that when applied to other models in LQG, the revised NJA works quite well in generating rotating metrics starting with their nonrotating seed metrics \citep{Liu:2020ola,Brahma:2020eos,Chen:2022nix}.

We also show that it is feasible, in principle, to constrain the LQG parameter $l$ using the Event Horizon Telescope (EHT) observed
shadows of the M87* and Sgr A* black holes. To be precise, we find that
the effects of parameter $l$ on the shadow size are more noteworthy than those on the the deviation of circularity
of the shadow silhouette. The Event Horizon Telescope Collaboration et al.  (\citeyear{EventHorizonTelescope:2019dse,EventHorizonTelescope:2019ggy,EventHorizonTelescope:2019jan,EventHorizonTelescope:2019pgp,EventHorizonTelescope:2019ths,EventHorizonTelescope:2019uob}) released the first horizon-scale image of the M87* black hole in 2019. Using a distance $d=16.8$ Mpc and estimated mass of M87* $M=(6.5 \pm 0.7) \times 10^9 M_\odot$  \citep{EventHorizonTelescope:2019dse,EventHorizonTelescope:2019pgp,EventHorizonTelescope:2019ggy}, the EHT results have bounds on the compact emission region size with angular diameter $\theta_d=42\pm 3\, \mu $as and circularity deviation $\Delta C\lesssim 0.1$ along with the central flux depression with a factor of $\gtrsim 10$ identified as the shadow.
The observed shadow image of the M87* black hole is compatible with the Kerr black hole as predicted by  GR. However, the present uncertainty in the measurement of spin and the relative deviation of quadrupole moments do not eliminate modification of Kerr black holes  \citep{EventHorizonTelescope:2019dse,EventHorizonTelescope:2019pgp,EventHorizonTelescope:2019ggy,Cardoso:2019rvt}.
In 2022, the EHT collaboration released the shadow results of black hole Sgr A* in the Milky Way depicting shadow angular diameter $d_{sh}= 48.7 \pm 7\,\mu$as and thick emission ring of
diameter $\theta_d=51.8\pm2.3\mu$as; considering a black hole of mass $M = 4.0^{+1.1}_{-0.6} \times 10^6 M_\odot $ and distance $8kpc$ from the Earth, the EHT images of Sgr A* are agreeing with the expected appearance of a Kerr black hole \citep{EventHorizonTelescope:2022exc,EventHorizonTelescope:2022urf,EventHorizonTelescope:2022vjs,EventHorizonTelescope:2022wok,EventHorizonTelescope:2022xnr,EventHorizonTelescope:2022xqj}. Furthermore, when compared with the EHT results for M87*, it exhibits consistency with the predictions of GR pushing across three orders of magnitude in central mass \citep{EventHorizonTelescope:2022xnr}.

The remainder of the paper is structured as follows. In Sect.~\ref{Sec-2}, we cover the geometric properties of the rotating polymerized black hole and the conformal diagrams. In Sect.~\ref{Sec-3}, we discuss the polymerized black hole shadows. Considering the observer at the equatorial plane, the  black hole shadow observables and their applications in estimating the LQG parameter and the black hole spin are presented in Sect.~\ref{Sec-4}. Constraints on the LQG parameter are derived from the EHT shadow bounds of Sgr A* and M87* for the inclination angles of $17^o$ and $50^o$ in Sect.~\ref{Sec-5}. We analyse our important findings and future prospects in Sect.~\ref{Sec-6}. 

\section{LQG-motivated  Polymerized Black Hole}\label{Sec-2}
A static and spherically symmetric black hole model in four-dimensional partially polymerized theory reads as  \citep{Peltola:2008pa,Peltola:2009jm}
\begin{eqnarray}\label{metric1}
ds^{2}&=&\left(\sqrt{1-\frac{l^2}{y^2}}-\frac{2M}{y}\right)dt^{2}-\frac{\left(1-\frac{l^2}{y^2}\right)^{-1}}{\left(\sqrt{1-\frac{l^2}{y^2}}-\frac{2M}{y}\right)}dy^{2} \nonumber\\
&-& y^2 (d\theta^{2}+\sin^{2}\theta d\phi^{2}),
\end{eqnarray}
with $M$ being the black hole mass and $l$ is the bounce radius. Equation (\ref{metric1}) has a unique  horizon at $y_H=\sqrt{4M^2+l^2}$  that demonstrates a quantum correction as a result of polymerization. In limited proper time, the solution evolves from the horizon at $y_H$ to the smallest radius $l$, and then expands  in infinite proper time to $y=\infty$ where the  solution is also asymptotically flat. All across, the Ricci and Kretschmann scalars are finite, nonsingular and rapidly dissipate away from the black hole. The polymerized black hole  (Equation (\ref{metric1})) thus characterizes a globally regular spacetime in a way that allows a sphere of radius $y=l$ to replace the curvature singularity at $y=0$ and bounce into an infinitely expanding Kantowski-Sachs spacetime. The solution although does violate the classical energy requirements, according to a calculation of the Einstein tensor, but the violations are of order $l^2/y^4 $ and thus diminish far beyond the bounce radius $l$ \citep{KumarWalia:2022ddq}.
This implies that the Schwarzschild solution which the Equation (\ref{metric1}) recovers in the limit $l \to 0$, is well approximated everywhere far outside the horizon, and that the exterior is equipped with nonzero quantum stress energy that is statistically insignificant for macroscopic black holes $(y_H \gg l)$. Additionally, the radial coordinate $y$ admits a minimum value of $y=l$, and the two-sphere at the center is known as the black bounce, whose geometries are represented by solutions to the Einstein equations with phantom scalar fields \citep{Bronnikov:2005gm,Bronnikov:2006fu,Bronnikov:2015kea,Bronnikov:2018vbs,Bronnikov:2021uta}.
\begin{figure*}
\begin{center}
	\includegraphics[scale=1.2]{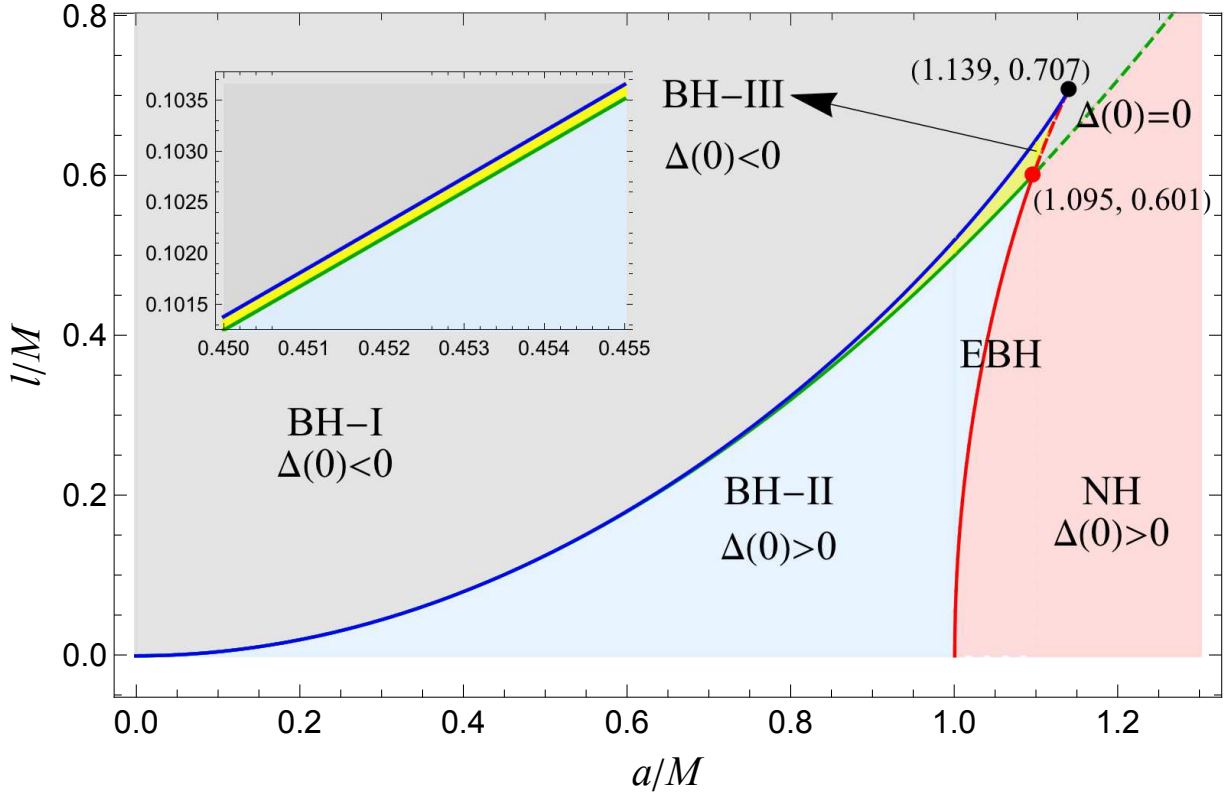}
	\caption{Parameter space ($a,l$) for LMRBH metric (\ref{metric3}). Acronyms are spelled in the text. The parameters along the red line describe an extremal black hole with degenerate horizons. The blue line is a transition surface, with one positive and two degenerate horizons where as on green line with  $\Delta(0)=0$,  for $l < 0.601M$ one has $\Delta(0)=0$ and two horizons while for $l > 0.601M$, we have $\Delta(0)=0$ only. The LMRBH extremal black hole refers to a black hole with $a>M$ \citep{Kumar:2022vfg}. }\label{parameter}
\end{center}	
\end{figure*}
The coordinate singularity at  $y=l$ in Equation (\ref{metric1}), can be explicitly eliminated by the coordinate transformation $y=\sqrt{l^2+r^2}$ resulting in the much simpler and regular equation which reads as
\begin{eqnarray}\label{metric2}
ds^{2}&=&\left(\frac{r-2M}{\sqrt{r^2+l^2}}\right)dt^{2}-\frac{1}{\left(\frac{r-2M}{\sqrt{r^2+l^2}}\right)}dr^{2}\nonumber\\
&-&(r^2+l^2)(d\theta^{2}+\sin^{2}\theta d\phi^{2}).
\end{eqnarray}
The radial coordinate $r$ assumes its entire range in this instance, i.e.,   $0\leq r\leq \infty$. Unlike Equation (\ref{metric1}), the horizon in this case is independent of $l$ and is always fixed at $2M$. The majority of the regular black holes that are closely related to a potential theory of quantum gravity might be seen as being inspired by quantum gravity rather than having been derived from it. As such, they constitute useful phenomenological models but make their physical justification less straightforward \citep{Ashtekar:2004eh}. However, by treating the quantum geometry corrections as an "effective" matter contribution, Walia (\citeyear{KumarWalia:2022ddq}) has shown that the phantom scalar field and the nonlinear electrodynamics field together sourced the static and spherically symmetric polymerized black hole (Equation (\ref{metric1})), and thus,  enhanced the significance of the polymerized black hole as an interesting solution to Einstein's field equations.  Equation (\ref{metric2}) also reduces to a Schwarzschild black hole in the $l \to 0$ limit.
\begin{figure*}[t] 
	\begin{centering}
		\begin{tabular}{p{9cm} p{9cm}}
		    \includegraphics[scale=0.75]{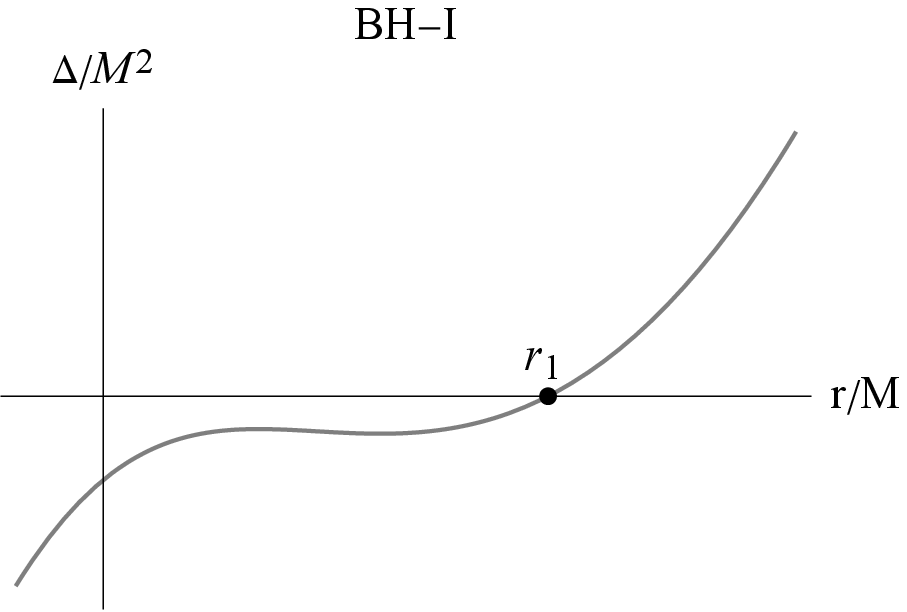}&
		    \includegraphics[scale=0.75]{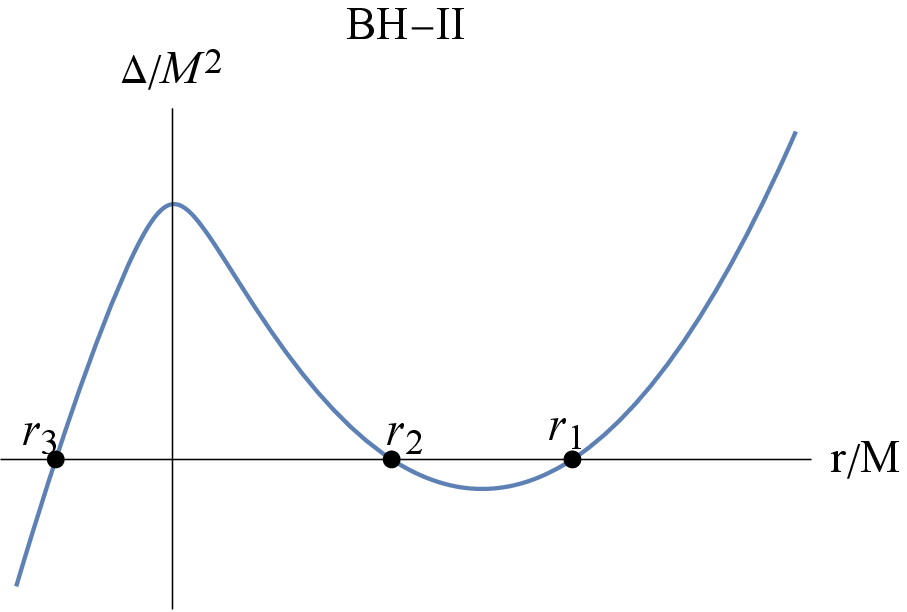}\\
		    \includegraphics[scale=0.4]{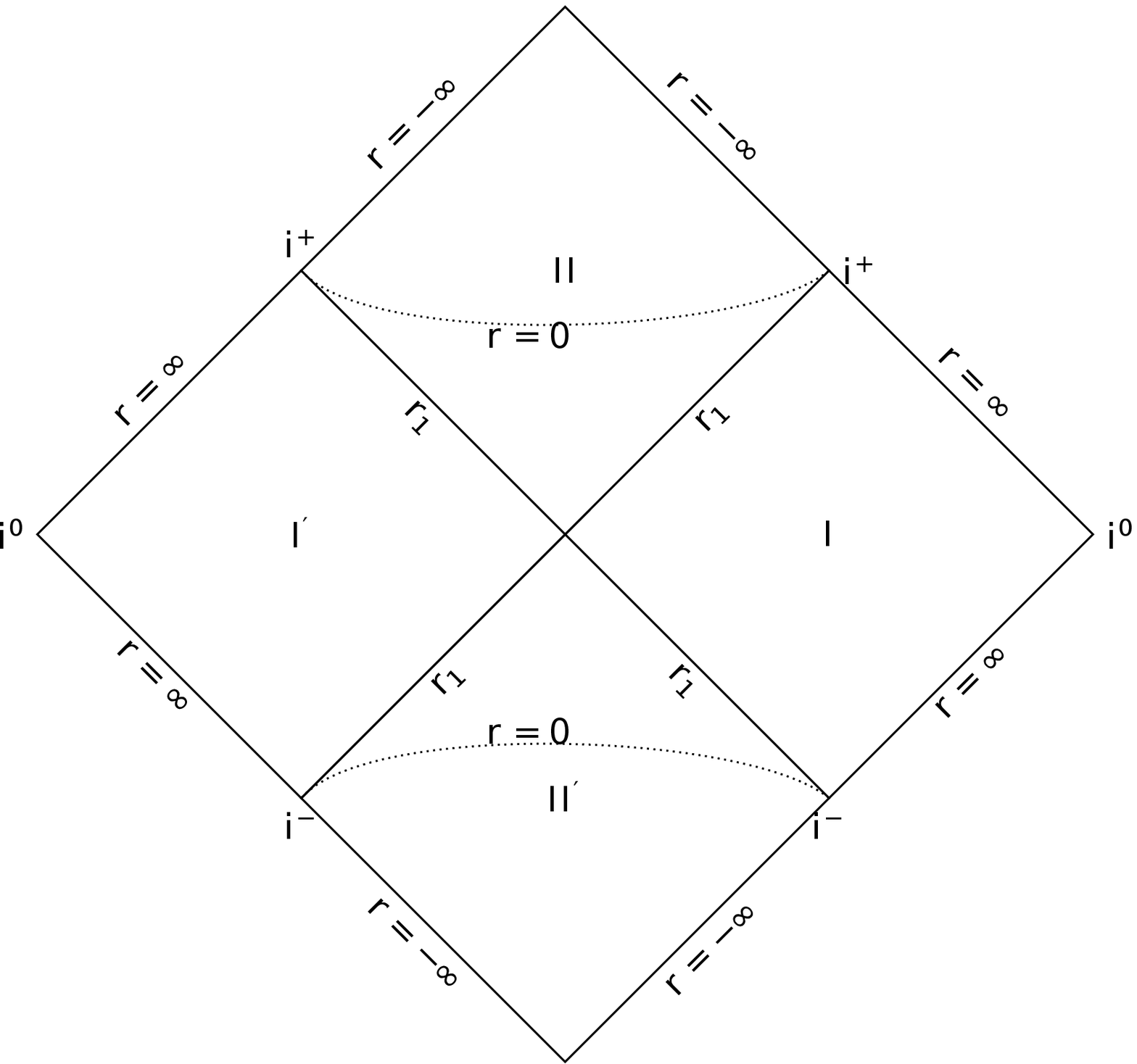}&
		     \includegraphics[scale=0.45]{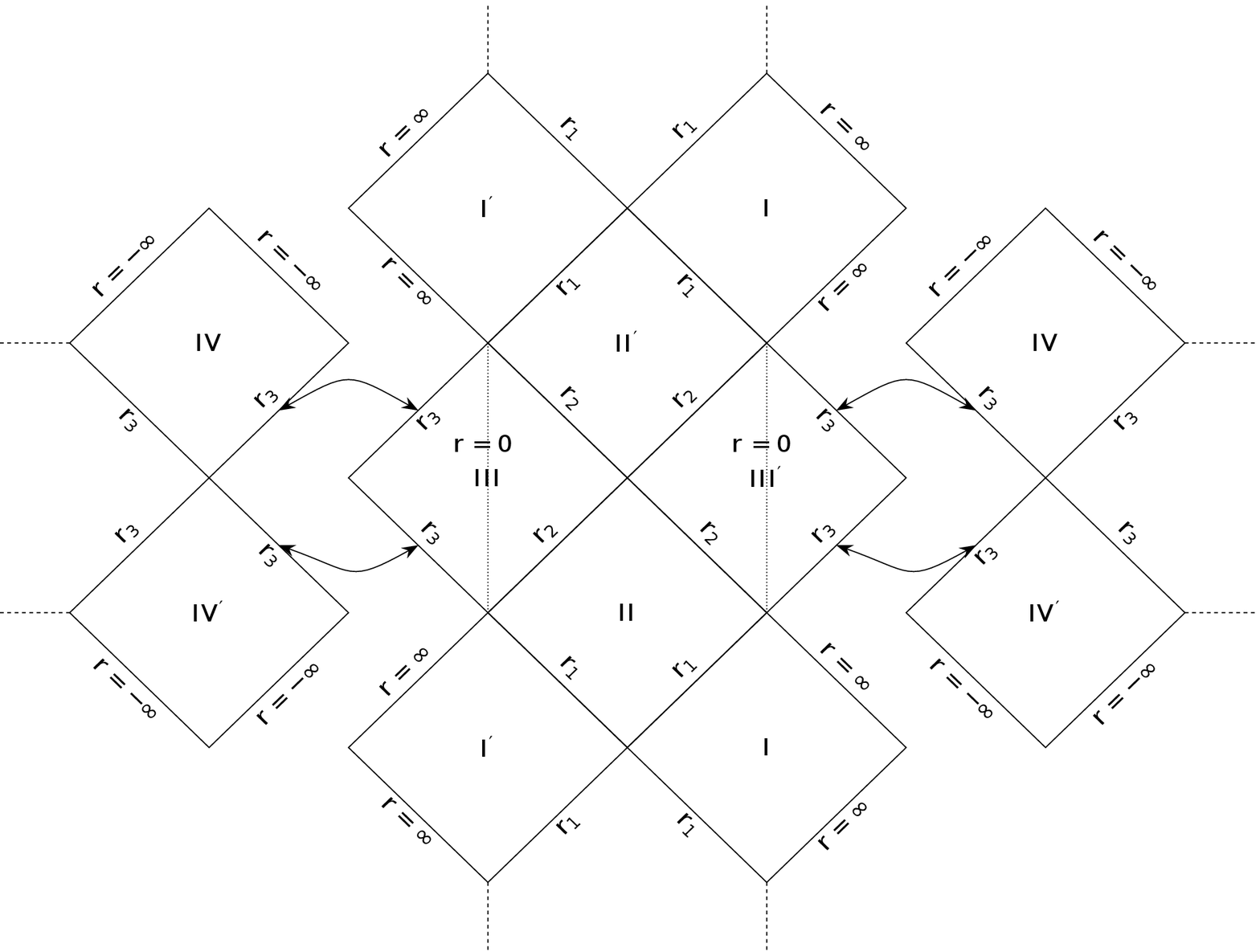}
			\end{tabular}
	\end{centering}
	\caption{Plot showing $\Delta(r)$ in the region of parameter space (Figure  \ref{parameter}) (a) BH-I, where $\Delta(r)=0$ admits only one positive root $r_1=$ corresponding to event horizon (Top left), (b) BH-II where $\Delta(r)=0$ admits two distinct positive roots $r_1$ (event horizon) and $r_2$ (Cauchy horizon), and one negative root $r_3$ corresponding to additional third horizon (Top right).
	Penrose-Carter diagrams of the parameter space $(M,a,l)$ showing the region (a) BH-I  (bottom left) and (b) BH-II (bottom right) \citep{Kumar:2022vfg}.}\label{pen4}
\end{figure*}

\subsection{Rotating black holes}
%The first  image of the black holes M87* and Sgr A* has been revealed by the EHT team \citep{EventHorizonTelescope:2019dse,EventHorizonTelescope:2022xnr}. These images are consistent with a Kerr black hole of general relativity, implying that it may not justify the existence of non-rotating black holes \citep{EventHorizonTelescope:2022xqj}.
The EHT observations of Sgr A* and M87* black hole shadows have confirmed that these astrophysical black holes are indeed rotating \citep{EventHorizonTelescope:2019dse,EventHorizonTelescope:2022xqj}. This serves as our motivation to search for an axisymmetric extension of Equation (\ref{metric2}), more specifically, an LMRBH that can be tested using the EHT observations. A direct loop quantization of stationary and axisymmetric black hole spacetime is still an unsolved problem. However, the  NJA introduced a revolutionary way to produce spinning spacetimes from a static, spherically symmetric seed metric without integrating any field metrics \citep{Newman:1965tw}. Additionally, the revised NJA was used to obtain the first-ever rotating black hole solution in the LQG \citep{Brahma:2020eos}, and has been subsequently used to produce rotating black holes in modified gravities \citep{Johannsen:2011dh,Ghosh:2014pba,Kumar:2017qws,Kumar:2020hgm,Kumar:2020owy}.
Starting with a partially polymerized static and spherically symmetric black hole solution (Equation (\ref{metric2})) and applying the revised NJA algorithm \citep{Azreg-Ainou:2014pra,Azreg-Ainou:2014aqa}, the resulting rotating spacetime LMRBH is always expressible in Boyer-Lindquist form, which is given as \citep{Kumar:2022vfg} 
\begin{eqnarray}\label{metric3}
ds^2 &=& \left(1-\frac{2M(r)\sqrt{r^2+l^2}}{\rho^2}\right) dt^2- \frac{\rho^2}{\Delta} dr^2 -\rho^2 d\theta^2  \nonumber\\ && + 4aM(r)\sqrt{r^2+l^2} \sin^2\theta dtd\phi- \frac{\mathbb{A}\sin^2\theta~}{\rho^2} d\phi^2
\end{eqnarray}
where
\begin{eqnarray}
M(r) &=& M-\frac{r-\sqrt{r^2+l^2}}{2} ~~~~~~ \rho^2 = r^2 + l^2 +a^2\cos^2\theta,\nonumber \\
\Delta &=& r^2+l^2 +a^2 -2 M(r) \sqrt{r^2+l^2},\nonumber \\
\mathbb{A} &=& (r^2+l^2 +a^2)^2-a^2 \Delta \sin^2\theta.
\end{eqnarray}
Although the rotating metric, derived from the NJA, does not guarantee that the LQG field equations can be satisfied, it captures key aspects of LQG, such as the global regularity of spacetime and the presence of a transition surface at the black hole center.  First, we note that LMRBH  (Equation (\ref{metric3})) encompasses the Kerr   spacetimes \citep{Kerr:1963ud} in the limit $l \to 0$, while in the limit $a \to 0$, the spherical LQG black hole (Equation (\ref{metric2})) is regained. Furthermore, when $a\to 0$, $M\to 0$, and $l\to 0$ the Equation (\ref{metric3}) gives flat spacetime.    

The null surface $\Delta(r)=0$, a coordinate singularity of the Equation (\ref{metric3}), defines the horizons of the LMRBH. It turns out that depending on the values of $a$ and $l$, $\Delta(r)=0$ can admit up to three real roots with one to three positive roots and can also have one negative root. We shall consider only positive roots as they correspond to the horizon. We identified the roots $r_1$, $r_2$, and  $r_3$ with $r_3 \le r_2 \le r_1$. The largest root, $r_1$, is always an event horizon, whereas $r_2$, if it exists, is always a Cauchy horizon. While $r_3$ is an additional horizon inside the Cauchy horizon and when $r_3<0$, it corresponds to a horizon inside the $r=0$ surface. The parameter space ($a,l$) for the LMRBH is depicted in  Figure ~\ref{parameter}. 
The spacetime structure of LMRBH spacetime intensely depends on the parameter $(M,\;a,\; l)$  as shown in  Figure ~\ref{parameter}. We have identified four regions: black holes with only one horizon (BH-I), black holes with an event  horizon and Cauchy horizon (BH-II), black holes with  three horizons (BH-III) and black holes with no horizon (NH). The boundaries shown by colored lines split these regions (see Kumar et al.  (\citeyear{Kumar:2022vfg}) for details), and lead to interesting geometrical features:
\begin{enumerate}
     \item \textbf{Region BH-I}: In this region of parameter space (gray region in Figure ~\ref{parameter}), $\Delta(r)=0$ has only one positive root corresponding to the single horizon.
    \item \textbf{Region BH-II}: Here (light blue region in Figure ~\ref{parameter}),  $\Delta(r)=0$ has two positive roots ($r_2,r_1$) $(r_2 < r_1)$ and one negative root ($r_3$). The former case corresponds to the Cauchy horizon ($r_2$) and event horizon ($r_1$).  This region is the most physically relevant and is referred to as the \emph{generic black hole}.
    \item \textbf{Region BH-III}: In this region (the yellow region in Figure ~\ref{parameter}), $\Delta(r)=0$ admits three positive roots, viz.,  $r_3$, $r_2$, and  $r_1$ corresponding to three horizons, viz., inner horizon ($r_3$), Cauchy horizon ($r_2$) and event horizon ($r_1$). They degenerate at the black dot as shown in Figure  \ref{parameter}.
    \item \textbf{Region NH}: Here (red region in Figure ~\ref{parameter}), $\Delta(r)=0$ has only one negative root. We refer to it as no-horizon spacetime, which, we show, is almost ruled out by EHT observations.
\end{enumerate}

Thus, unlike the Kerr black hole, $a = M$ does not yield an extremal black hole in the LMRBH. It turns out that  the LMRBH can admit one negative root and one to three positive roots. Thus, the LMRBH, unlike Kerr's black hole, admits up to three horizons. The case of two distinct positive horizons, viz., $r_1, r_2\;  (r_1>r_2)$ is referred to as a \textit{generic black hole} with  the event horizon ($r_1$) and the Cauchy horizon ($r_2$). For further details see Ref. Kumar et al. (\citeyear{Kumar:2022vfg}) where we have given horizon structure and conformal diagrams of LMRBH. Thus, we have an LMRBH that is  regular everywhere and asymptotically encompasses the Kerr black hole as a particular case. Also, the LMRBH metric (Equation \ref{metric3})  describes a multi-horizon rotating black hole in the sense that it can admit up to three horizons, and that an extremal LMRBH, unlike the Kerr black hole, refers to a black hole with angular momentum $a>M$. Besides having properties, viz., asymptotic flatness and regularity, 
next, we check the separability of the geodesic equations, which helps us to test the LMRBH spacetime with its shadow via EHT observations. 

\paragraph{Penrose diagram:}
The Penrose diagram of  region BH-I (see Figure\ref{pen4} (left)) is precisely the same as  its spherical counterpart \citep{Peltola:2008pa}.  It reveals two exterior regions, the black hole and the white hole interior regions, and two quantum-corrected interior regions \citep{Peltola:2008pa}. The transition surface, $r=0$ is space-like and hidden behind the event horizon. Unlike Kerr black holes, the Penrose diagram of the generic black hole region BH-II has quantum-corrected regions ($r_3$ to $r=-\infty$) where  $t$ and $r$  swap  roles and replicate infinite times horizontally in both paths. The time-like transition surface substitutes the classical ring singularity occurring in the Kerr black hole at $r=0, \theta =\pi/2$ ( see Figure \ref{pen4} (right)). Region BH-III has a similar Penrose diagram and hence is not presented \citep{Kumar:2022vfg}. 

\section{Black hole shadow}\label{Sec-3}
The null geodesics describing the photon orbits around the black hole are intriguing because of their observational significance in analyzing the gravitational impact of the black holes on the surrounding radiation. Light rays, coming from the background source behind the black hole, with an impact parameter greater than the critical value, get strongly deflected around the black hole and reach the observer, whereas those with an impact parameter smaller than the critical value fall into the event horizon, resulting in a dark region on the observer's sky, \emph{shadow}, encompassed by the bright photon ring \citep{Synge:1966,bardeen1973,Luminet:1979,Cunningham:1973}. The influential work led by Synge (\citeyear{Synge:1966}) and Luminet (\citeyear{Luminet:1979}), who furnished the formula to
measure the angular radius of the photon, captured the region around the Schwarzschild black hole by identifying the diverging light deflection angle. Later, in the pioneering work Bardeen (\citeyear{bardeen1973}), the shadow of the Kerr black hole was investigated  and showed that the spin  caused a distortion in shadow shape. The photon ring, encompassing the black hole shadow, explicitly depends on the spacetime geometry and thus its shape and size is a potential tool to determine the black hole parameters
and to reveal the valuable information regarding the near-horizon field features of gravity. Later, a flurry of activities in the analytical/numerical investigations and observational studies of shadows for a large variety of black holes have been reported \citep{Vries:2000,Shen:2005cw,Amarilla:2010zq,Yumoto:2012kz,Amarilla:2013sj,Atamurotov:2013sca,Abdujabbarov:2016hnw,Abdujabbarov:2015xqa,Cunha:2018acu,Mizuno:2018lxz,Mishra:2019trb,Shaikh:2019fpu,Kumar:2020yem}. Shadows have also been investigated for black hole parameter estimations \citep{Kumar:2018ple} and testing theories of gravity \citep{Kramer:2004hd}. 

We start with the Hamilton-Jacobi equation to determine the null geodesics followed by photons in the LMRBH spacetime (Equation \ref{metric3}) \citep{Carter:1968rr,Chandrasekhar:1985kt}  
\begin{eqnarray}
\label{HmaJam}
\frac{\partial S}{\partial \tau} = -\frac{1}{2}g^{\alpha\beta}\frac{\partial S}{\partial x^\alpha}\frac{\partial S}{\partial x^\beta} ,
\end{eqnarray}
where $\tau$ is the affine parameter along the geodesics, and $S$ is the Jacobi action.  Equation (\ref{metric3}) is a time translational and rotational
invariant, which leads to conserved quantities along null geodesics,
namely, energy $\mathcal{E}=-p_t$ and axial angular momentum $\mathcal{L}=p_{\phi}$, where $p_{\mu}$ is the photon's four momentum. In addition, the Petrov type D
character of Equation (\ref{metric3}) ensures the existence of Carter's
separable constant. Therefore, the Jacobi action can be written as
\begin{eqnarray}
S=-{\cal E} t +{\cal L} \phi +S_r(r)+S_\theta(\theta) \label{action},
\end{eqnarray}
where $S_r(r)$ and $S_{\theta}(\theta)$, respectively, are only functions  of the $r$ and $\theta$ coordinates. Therefore, the four integrals of motions, the Lagrangian, energy $\cal E$, axial angular momentum $\cal L$, and the Carter constant associated with the latitudinal motion of the photon, are sufficient to determine the geodesic equations of motion in the first-order differential form \citep{Carter:1968rr,Chandrasekhar:1985kt} as follows:
\begin{eqnarray}
\Sigma \frac{dt}{d\tau}&=&\frac{r^2+l^2+a^2}{\Delta}\left({\cal E}(r^2+a^2)-a{\cal L}\right)  \nonumber\\ &&-a(a{\cal E}\sin^2\theta-{\mathcal {L}}),~ \label{tuch}\\
\Sigma \frac{dr}{d\tau}&=&\pm\sqrt{\mathcal{R}(r)}\ ,\label{r}\\
\Sigma \frac{d\theta}{d\tau}&=&\pm\sqrt{\Theta(\theta)}\ ,\label{th}\\
\Sigma \frac{d\phi}{d\tau}&=&\frac{a}{\Delta}\left({\cal E}(r^2+l^2+a^2)-a{\cal L}\right)-\left(a{\cal E}-\frac{{\cal L}}{\sin^2\theta}\right),\label{phiuch}
\end{eqnarray}
where  $\mathcal{R}(r)$ and $\Theta(\theta)$, respectively, pertain to the following radial and polar motion effective potentials: 
\begin{eqnarray}\label{06}
\mathcal{R}(r)&=&\left[(r^2+l^2+a^2){\cal E}-a{\cal L}\right]^2-\Delta[{\cal K}+(a{\cal E}-{\cal L})^2]\label{rpot},\quad \\ 
\Theta(\theta)&=&{\cal K}-\left[\frac{{\cal L}^2}{\sin^2\theta}-a^2 {\cal E}^2\right]\cos^2\theta.\label{theta0}
\end{eqnarray}
where the separability constant $\mathcal{K}$ is related to the Carter constant $\mathcal{Q}$ through $\mathcal{Q}=\mathcal{K}+(a\mathcal{E}-\mathcal{L})^2$ \citep{Carter:1968rr,Chandrasekhar:1985kt}, which, in essence, represents the isometry of Equation (\ref{metric3})  along the second-order Killing tensor field. The photon's $\theta$-motion is influenced by the $\mathcal{Q}$, but restricted  to an equatorial plane when $\mathcal{Q}=0$. Meanwhile, the $\mathcal{L}$ governs the $\phi$ motion. Contrary to the Schwarzschild black hole, where all null circular orbits are planar due to spherical symmetry, i.e., orbits with $\Dot{\theta}= 0$, the rotating black hole also has nonplanar orbits because of the effects of frame dragging. 
\begin{figure*}
    \begin{tabular}{p{8cm} p{8cm}}
   \includegraphics[scale=0.85]{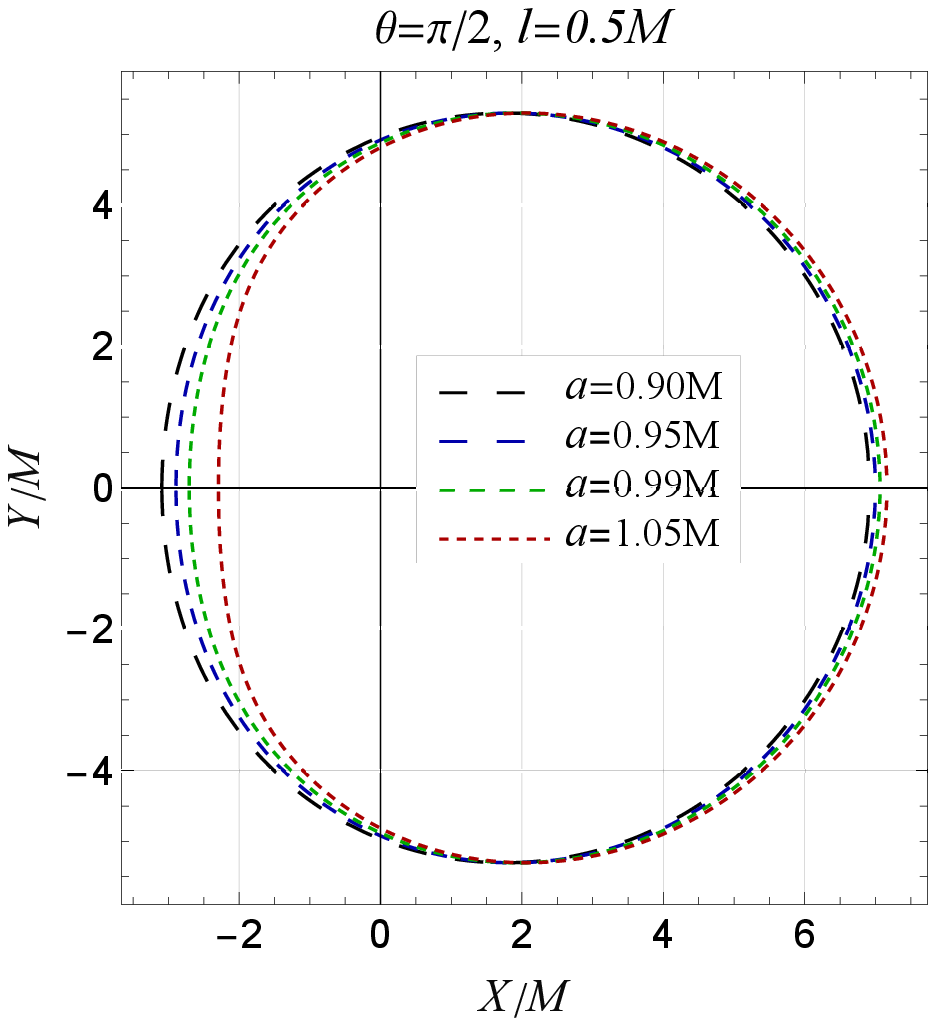}&
    \includegraphics[scale=0.85]{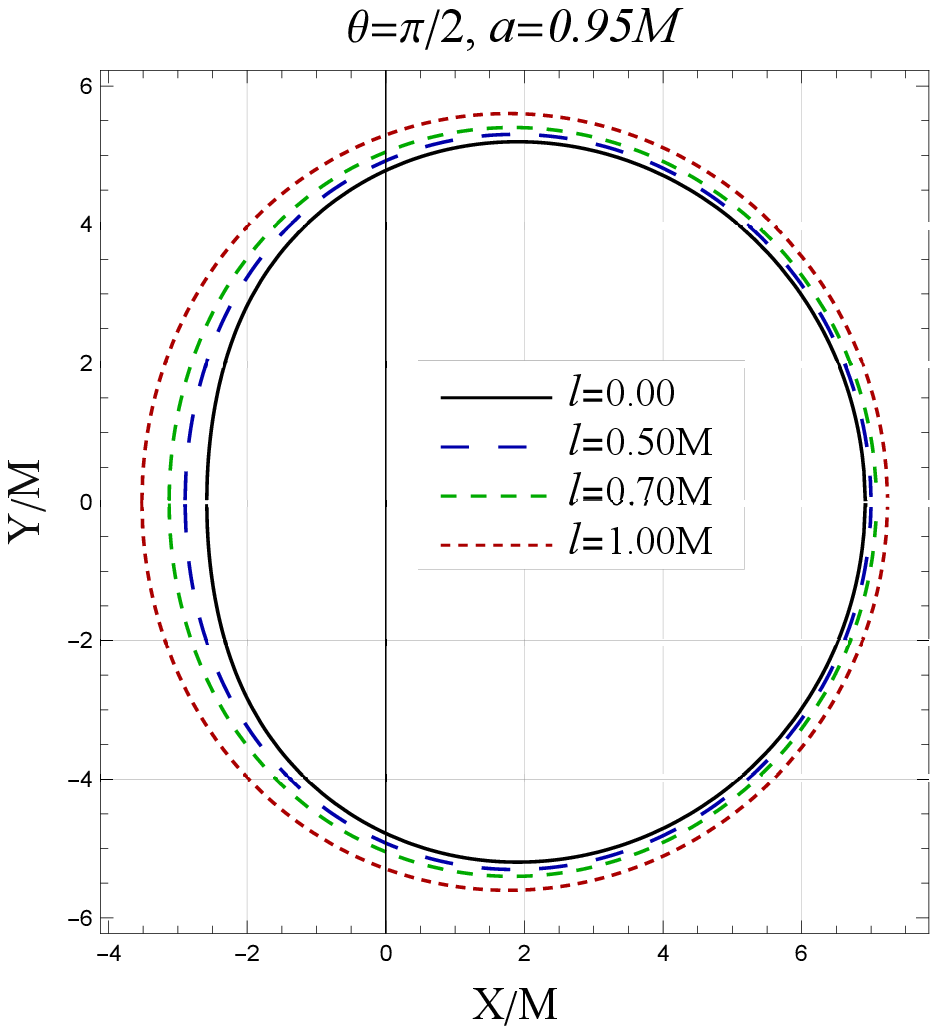}
    \end{tabular}
    \caption{Shadow silhouette of the LMRBH for $l=0.5$ with varying $a$  (\textit{left}) and for $a=0.95$ with varying $l$ (\textit{right}) as seen from the equatorial plane, i.e., inclination angle $\theta_o=\pi/2$.}\label{fig2}
    \begin{tabular}{p{8cm} p{8cm}}
     \includegraphics[scale=0.87]{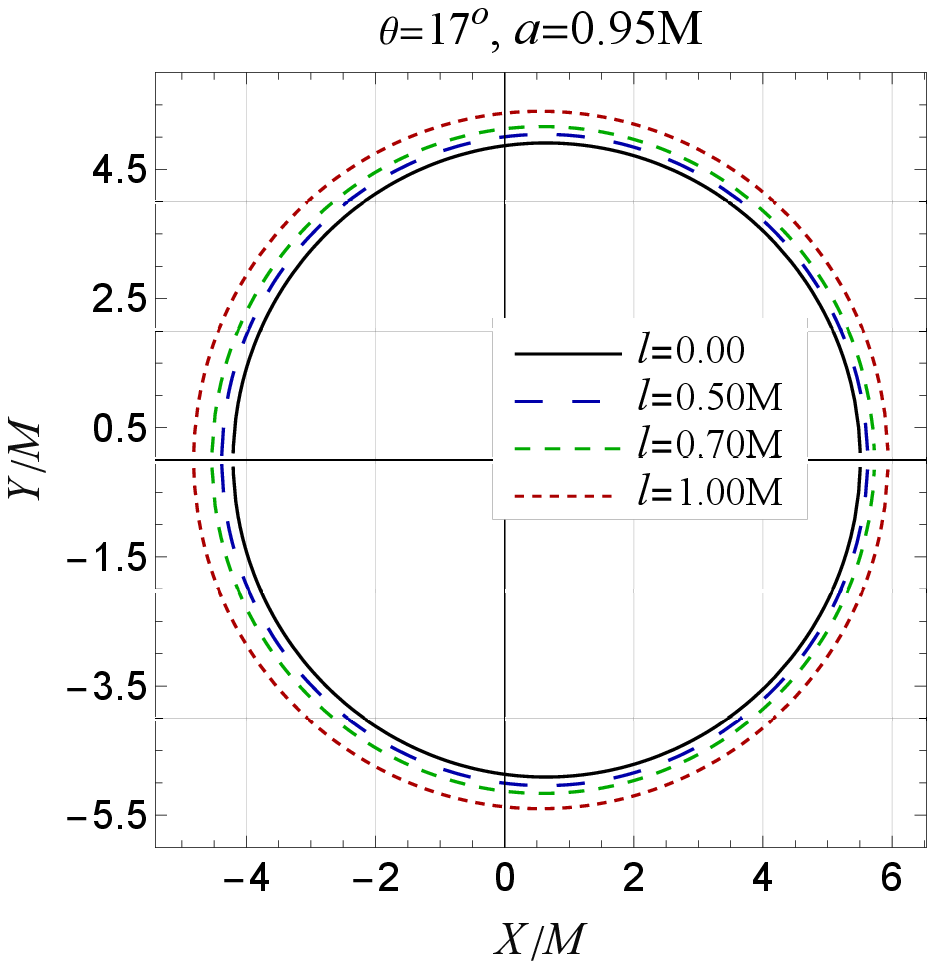}& 
    \includegraphics[scale=0.85]{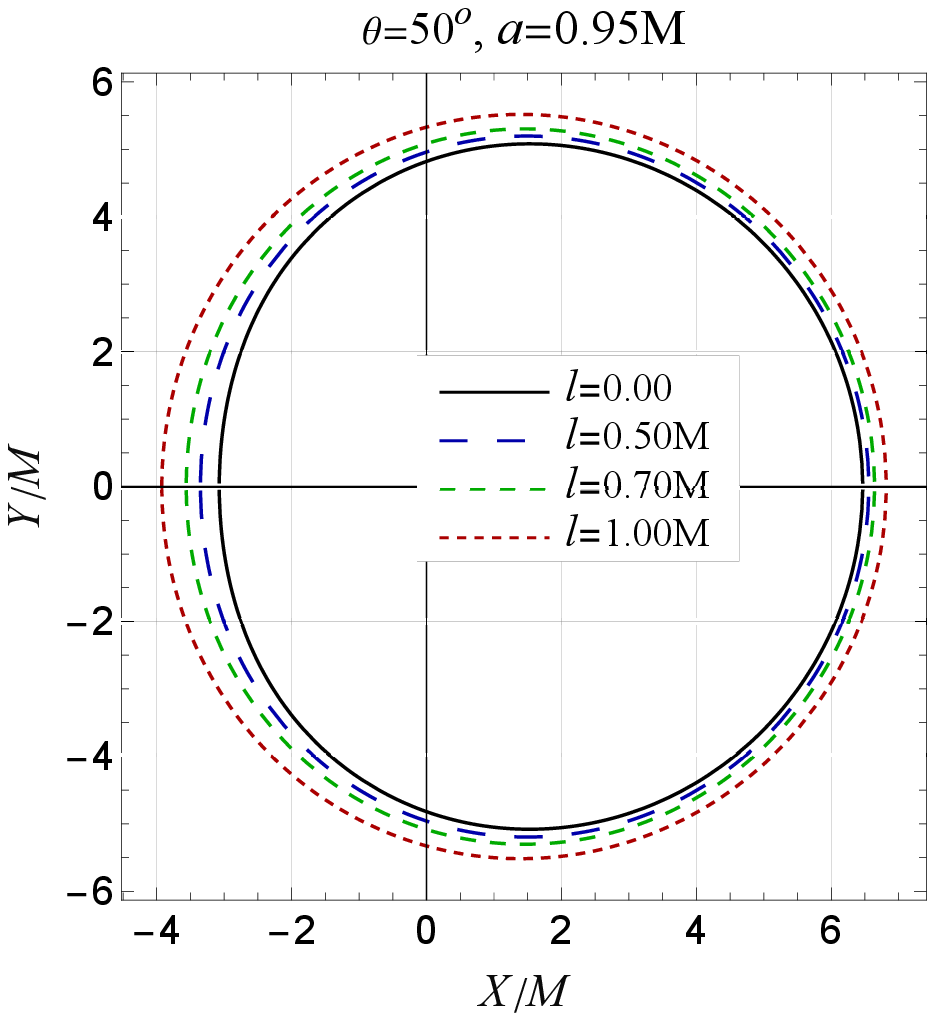}
    \end{tabular}
    \caption{Shadow silhouette of the LMRBH for $a=0.95$ with varying $l$ at inclination angle (\textit{left}) $\theta_o=17\degree$ and (\textit{right}) $\theta_o=50\degree$.}\label{fig3}
\end{figure*}
The black hole shadow silhouette  is outlined by the unstable spherical photon orbits, which can be determined by solving $\dot{r}=\ddot{r}=0$ from Eqs.~(\ref{r}) and (\ref{06}). The radii $r_p$ of the photon spherical orbits is positive root of the following equations:
\begin{equation}
\mathcal{R}|_{r=r_p}=\left.\frac{\partial \mathcal{R}}{\partial r}\right|_{r=r_p}=0,\,\, \text{and}\,\, \left.\frac{\partial^2 \mathcal{R}}{\partial r^2}\right|_{r=r_p}> 0.\label{vr} 
\end{equation}
To proceed further, following Chandershaker (\citeyear{Chandrasekhar:1985kt}), we can introduce the dimensionless parameters  
$\xi\equiv \mathcal{L}/\mathcal{E},\qquad \eta\equiv \mathcal{K}/\mathcal{E}^2$ to reduce the degree of freedom of photons geodesics to one.  Solving Equation~(\ref{vr}) for Equation~(\ref{rpot}) results in the critical impact parameters as follows: \citep{Chandrasekhar:1985kt}
\begin{eqnarray}\label{impactparameter}
\xi_c &=& \frac{(a^2 +l^2+r_p^2) \Delta '(r_p)-4 r_p \Delta (r_p)}{a \Delta '(r_p)}\nonumber\\
\eta_c &=& \frac{1}{a^2 \Delta '(r_p)^2}(16r_p^2\Delta (r_p) (a^2-\Delta (r_p)) 
\nonumber\\ &&+ (l^2+r_p^2)\Delta'(r_p)\left[ 8 r_p \Delta(r_p)-(l^2+r_p^2)\Delta'(r_p)\right])~~~~~
\end{eqnarray}
where $'$ stands for the derivative with respect to the radial coordinate.  Equation~(\ref{impactparameter}) in the limit $l \to 0$ reduces to the following:  
\begin{eqnarray}
\xi_c^K &=& \frac{a^2 (M + r_p) + r_c(r_p-3M)}{a (M- r_p)}  \nonumber\\
\eta_c^K &=&  \frac{r_p^3 \left[4 a^2 M - r_p(r_p-3M)^2\right]}{a^2 \left(M-r_p\right)^2}
\end{eqnarray}
which are exactly the same as that for the Kerr black hole \citep{Chandrasekhar:1985kt}.  For light rays coming from the bright source, there are three possible trajectories (i) capture orbit, (ii) scatter orbit, and (iii) unstable orbit. The light rays, which are plunged into the black hole, form the black hole shadow. The shape of the black hole shadow depends on the spin $a$ and observation angle $\theta_0$ with respect to the spin axis. 
The relationship between the observer's celestial coordinates, $X$ and $Y$, and two constants, $\xi_c$ and $\eta_c$, is derived using the tetrad components of the four momentum $p^{(\mu)}$ and geodesic Eqs.~(\ref{tuch}),  (\ref{r}), (\ref{th}), and (\ref{phiuch}) as 
\begin{eqnarray}
&&X= -r_o\frac{p^{(\phi)}}{p^{(t)}} = -\left. r_o\frac{\xi_c}{\sqrt{g_{\phi\phi}}(\zeta-\gamma\xi_c)}\right|_{(r_o,\theta_o)},\nonumber\\
&&Y = r_o\frac{p^{(\theta)}}{p^{(t)}} =\pm\left. r_o\frac{\sqrt{\Theta(\theta)}
}{\sqrt{g_{\theta\theta}}(\zeta-\gamma\xi_c)}\right|_{(r_o,\theta_o)},~\label{Celestial}
\end{eqnarray} 
where 
\begin{eqnarray}
\zeta=\sqrt{\frac{g_{\phi\phi}}{g_{t\phi}^2-g_{tt}g_{\phi\phi}}},\qquad \gamma=-\frac{g_{t\phi}}{g_{\phi\phi}}\zeta.
\end{eqnarray}
where the coordinates $X$ and $Y$ in Eq.~(\ref{Celestial}), respectively, are  the apparent displacement along the perpendicular and parallel axes to the projected axis of the black hole symmetry. Thus, an observer can see the stereographic projection of black hole shadow defined by the celestial coordinates established by Bardeen (\citeyear{bardeen1973}) at radial infinity ($r_o\to\infty$) and an inclination angle $\theta_0$, as
\begin{equation}\label{pt}
X=-\xi_c\csc\theta_o,\qquad Y=\pm\sqrt{\eta_c+a^2\cos^2\theta_o-\xi_c^2\cot^2\theta_o}.
\end{equation} 
For an observer in the equatorial plane ($\theta_0=\pi/2$), Eq.~(\ref{pt}), reduces to
\begin{equation}
    \{X,Y\}=\{-\xi_c,\pm\sqrt{\eta_c}\}
\end{equation}
The parametric plots of Equation~(\ref{pt}) in the ($X, Y$) plane cast a variety of black hole shadows for different ranges of parameters as shown in figures.~\ref{fig2} and ~\ref{fig3}. For $a=l=0$, the contour of a Schwarzschild  black hole shadow from Eq.~(\ref{pt}) takes the form $X^2+Y^2=27M^2$, which ensures that the shadow is perfectly circular. The shadow of nonrotating LQG black holes, when compared with the Schwarzschild black holes is larger    
and the size of the shadow increases with parameter $l$ \citep{KumarWalia:2022ddq}. We depict the shadow silhouette of the LMRBH spacetime for various parameter values in Figure  \ref{fig3}. In fact, the shadow size, for a fixed value of spin $a$, increases and becomes more distorted with increasing $l$.  Additionally, we also see a horizontal shift in the shadow along the $X$-axis, with an increase in inclination angle $\theta_0$ and the spin $a$. 
\begin{figure}
\begin{center}
	\includegraphics[scale=0.85]{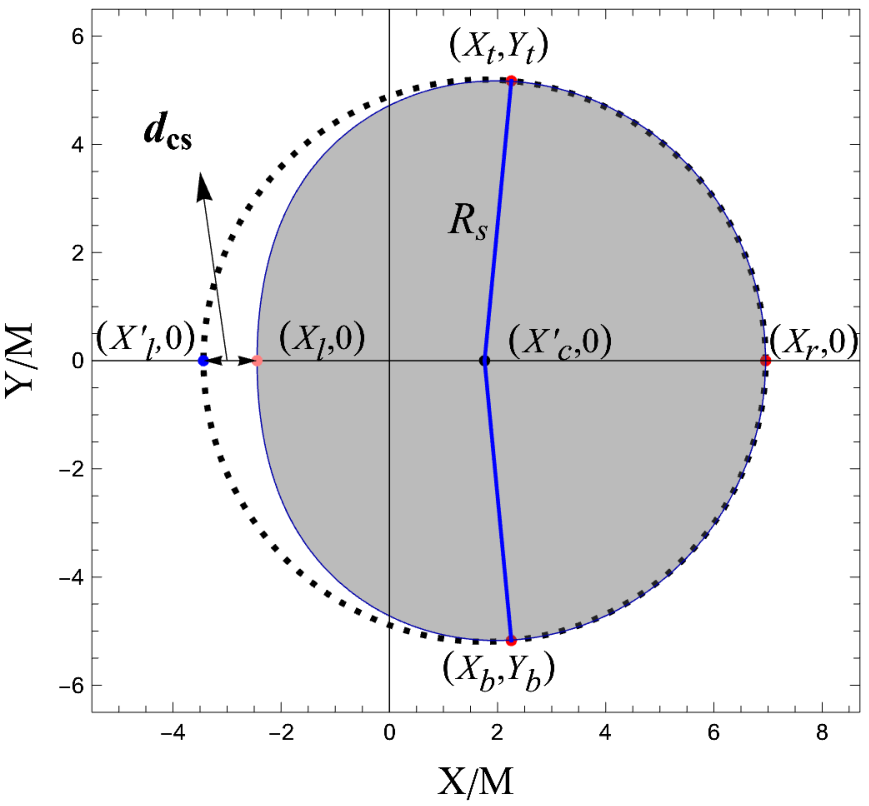}
	\caption{The shadow observables, radius $R_s$ and distortion $\delta_s=d_{cs}/R_s$, for the apparent  shape of shadow of the LMRBH ($a=0.99M$, $l=0.4M$ and $\theta_0=\pi/2$). The distortion is the difference between the left endpoints of the dotted circle and of the shadow. }\label{dis}
\end{center}
\end{figure}
\section{Observables and Estimation of black hole parameter }\label{Sec-4}
We aim to properly understand the LMRBH model. Hence, using shadow observables, we also estimate the parameters associated with LMRBH assuming that the observer is in the equatorial plane. It may be beneficial to ultimately determine information about an LMRBH.

We will employ two independent methods, Hioki-Maeda (\citeyear{Hioki:2009na}) and Kumar-Ghosh (\citeyear{Kumar:2018ple}), to characterize the LMRBH shadows using the shadow observables.  Prior knowledge of these observables through observations can be used to back estimate the black hole parameters. We assume the observer is in the equatorial plane, i.e., at an inclination angle $\theta_0=\pi/2$.

\begin{figure}
\begin{center}
	\includegraphics[scale=0.65]{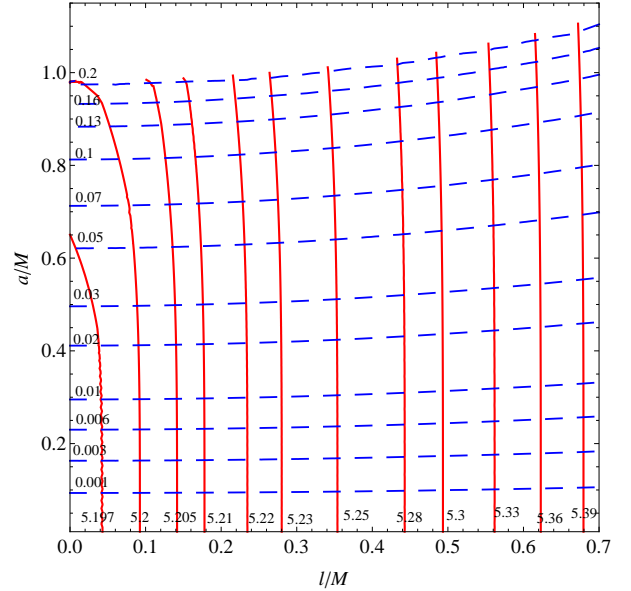}
	\caption{Contour plot of the shadow observables $R_s/M$ (red solid lines) and $\delta_s$ (blue dashed lines). The given lines of $R_s$ and $\delta_s$ intersect at a unique point ($a, l$) in the parameter space plane. }\label{contour1}
\end{center}
\end{figure}

\paragraph{Hioki-Maeda method:} Hioki and Maeda (\citeyear{Hioki:2009na})  proposed  two observables, radius $R_s$ and distortion $\delta_s$, to characterize the black hole shadow silhouette. They used a reference perfect circle with a center $(X'_c, 0)$ that coincides with the shadow silhouette at three points, $(X_t,Y_t)$,  $(X_b,Y_b)$,  $(X_r,0)$,  to approximate the shape of the black hole shadow  as shown in Figure ~\ref{dis}. The points $(X_l,0)$,  $(X'_l,0)$, represent the intersections of the shadow silhouette and reference circle with the horizontal axis $(Y=0)$, respectively, such that $d_{cs}=|X'_l-X_l|$ determines the potential dent on the black hole shadow in the direction perpendicular to the black hole rotational axis (see Figure ~\ref{dis}). We can identify the radius of the shadow as the radius of the reference circle $R_s$  and the distortion $\delta_s=d_{cs}/R_s$. 
These observables are defined as follows \citep{Hioki:2009na}:
\begin{eqnarray}
R_s &=& \frac{(X_t-X_r)^2+Y_t^2}{2|X_r-X_t|},\nonumber\\
\delta_s &=& \frac{|X'_l-X_l|}{R_s}.
\end{eqnarray}
We determine the black hole parameters by combining two contour plots for these observables, resulting in a one-to-one correspondence between ($a$, $l$) and ($R_s$, $\delta_s$).
With the prior knowledge of radius $R_s$ and distortion $\delta_s$, we can accurately calculate the spin parameter $a$ and the parameter $l$ using Figure  \ref{contour1} and Table \ref{table1}. 

\begin{table}
	\centering	
	\begin{tabular}
	{ |p{1.6cm}|p{1.6cm}|p{1.6cm}|p{1.6cm}| }
		\hline
		$R_s/M$ &  $\delta_s$ & $a/M$ & $l/M$  \\
		\hline\hline
5.197 & 0.01 & 0.2884 & 0.0437 \\
 \hline
5.197 & 0.02 & 0.4038 & 0.0405 \\
 \hline
5.21 & 0.01 & 0.2897 & 0.1788 \\
 \hline
5.21 & 0.10 & 0.8131 & 0.1684 \\
\hline
5.23 & 0.02 & 0.4129 & 0.2803 \\
\hline 
5.23 & 0.20 & 0.9808 & 0.2673 \\
\hline 
5.30 & 0.01 & 0.3058 & 0.4934 \\
\hline
5.30 & 0.20 & 1.023 & 0.4866 \\
\hline
5.39 & 0.05 & 0.6787 & 0.6868 \\
\hline 
5.39 & 0.16 & 1.038 & 0.6735 \\
\hline
5.39 & 0.20 & 1.078 & 0.6728 \\
\hline
\end{tabular}
\caption{Estimated values of parameters $a/M$ and $l/M$ from given shadow observables $R_s$ and $\delta_s$.}\label{table1}
\end{table}

\begin{figure}
\begin{center}
	\includegraphics[scale=0.65]{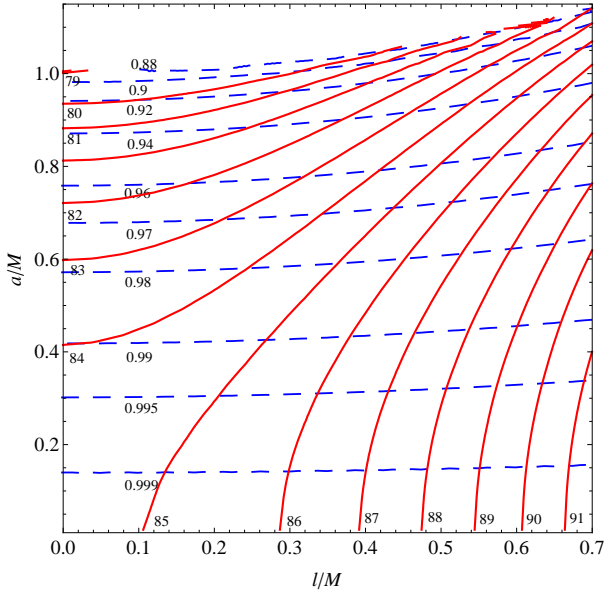}
	\caption{Contour plot of the observables $A/M^2$ (red solid lines) and $D$ (blue dashed lines). The given lines of $A$ and $D$ intersect at a unique point ($a, l$) in the parameter space plane. }\label{contour2}
\end{center}
\end{figure}

\paragraph{Kumar-Ghosh method:}
Kumar and Ghosh (\citeyear{Kumar:2018ple}) proposed an alternative method to characterize the black hole shadows using the observable, namely, area ($A$) enclosed by a black hole shadow and the shadow oblateness ($D$). The potential advantages of using these observables are that they can be used coordinate independently  by different teams analyzing the noisy observational data, and they do not require  comparing the shadow silhouette with a perfect circle and are thus applicable to a general shadow of any shape and size. 
The observable $A$  is defined by 
\begin{equation}\label{Area1}
A=2\int{Y(r_p) d X(r_p)}=2\int_{r_p^{-}}^{r_p^+}\left( Y(r_p) \frac{dX(r_p)}{dr_p}\right)dr_p,
\end{equation}     
By defining the dimensionless observable $D$, the ratio of the horizontal and vertical diameters as the oblateness \citep{Takahashi:2004xh, Grenzebach:2015oea,Tsupko:2017rdo}, a degree of distortion (circular asymmetry), a characterization of the rotating black hole's shadow can be done. The
oblateness can be as follows:
\begin{equation}\label{Oblateness}
D=\frac{X_r-X_l}{Y_t-Y_b}.
\end{equation}
The subscripts $r, l, t$, and $b$, respectively,  stand for the right, left, top, and bottom of the shadow silhouette. $D=1$ for a spherically symmetric black hole shadow, but $\sqrt{3}/2\leq D< 1$ for a Kerr shadow \citep{Tsupko:2017rdo}. The shadow structure (see Figure ~\ref{fig3}) clearly reveals that the parameters $a$ and $l$ have a substantial effect on both the shadow area and  oblateness. It is worth noting that a single shadow observable, either $A$ or $D$, will lead to degeneracy in estimating more than one black hole parameters. However, a set of shadow observables, ($A, D$) can uniquely determine two parameters. A one-to-one correspondence between the observables ($A, D$) and parameters ($a, l$) is depicted listed in Table \ref{table2} and shown in Figure ~\ref{contour2}.  

\begin{table}
	\centering	
	\begin{tabular}
	{ |p{1.6cm}|p{1.6cm}|p{1.6cm}|p{1.6cm}| }
		\hline
		$A/M^2$ &  $D$ & $a/M$ & $l/M$  \\
		\hline\hline
 79 & 0.92 & 0.9363 & 0.0955 \\
 \hline
 80 & 0.9 & 1.017 & 0.4292 \\
 \hline
 83 & 0.97 & 0.6803 & 0.2147 \\
 \hline
 83 & 0.92 & 1.009 & 0.5655 \\
 \hline
 85 & 0.99 & 0.4168 & 0.2706 \\
 \hline
 85 & 0.999 & 0.1341 & 0.1367 \\
 \hline
 86 & 0.999 & 0.1368 & 0.1329 \\
 \hline
 86 & 0.97 & 0.7184 & 0.5167 \\
 \hline
 87 & 0.96 & 0.8273 & 0.6358 \\
 \hline
 87 & 0.995 & 0.3096 & 0.4338 \\
 \hline
 90 & 0.995 & 0.3255 & 0.6348 \\
 \hline
 91 & 0.999 & 0.1456 & 0.6791 \\
 \hline
\end{tabular}
\caption{Estimated values of parameters $a/M$ and $l/M$.}\label{table2}
\end{table}
\section{Constraints from the EHT Observation}\label{Sec-5}
In spite of several astrophysical phenomena, including accretion flow, jet outflow, gravitational lensing, emission phenomena, etc., the black hole shadow shape acts as a direct diagnostic test of a strong-field gravity since it is the most obvious manifestation of the background spacetime. The black hole shadow boundary is constructed by the photons that can go the closest to the black hole but still manage to escape the black hole gravitational field and reach the observer, as a result of the shadow bearing imprints of the strong-field  characteristics  \citep{Jaroszynski:1997bw,Falcke:1999pj}, so much so that the  black hole shadow observations by the EHT collaboration have opened up an exciting arena to make a precision test of the gravitational theory in the strong and relativistic field regimes. It is worth noting that the current angular resolution of the EHT is not enough to capture the imprints of the  quantum gravity effects. The EHT collaboration \citep{EventHorizonTelescope:2022xnr}, a very long baseline interferometry experiment, that measures radio brightness distributions at a wavelength of 1.3 mm on the sky with an unprecedented angular resolution of $20\mu$as, has captured the horizon-scaled emission image of the Sgr A* and M87* black holes. Both shadows have some common features; the shadow center has a significant depression of brightness due to photon capture by the event horizon of the black hole, which is enclosed by a bright nearly circular ring. The  EHT collaboration put bounds on the size and the shape of this ring. Using the EHT observations of the shadows cast by black holes M87* and Sgr A*, we explore potential constraints on the LMRBH parameters. 
\begin{figure}
\begin{center}
	\includegraphics[scale=0.75]{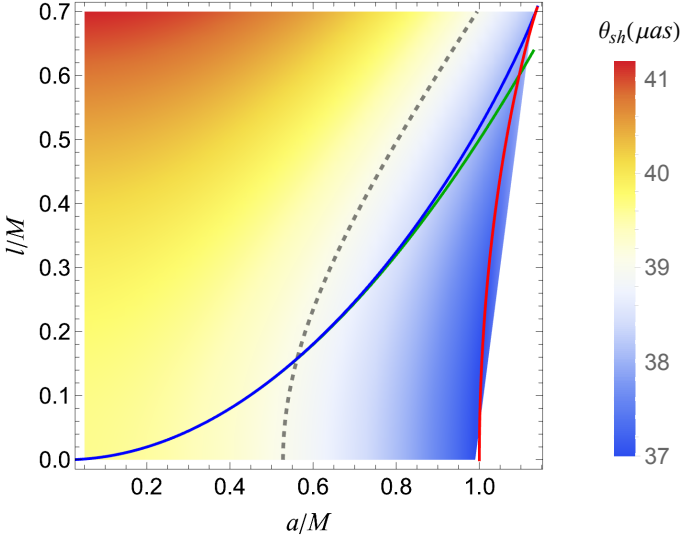}
	\caption{LMRBH shadow angular diameter $\theta_{sh}$, in units of $\mu$as, as a function of ($a, l$) for an inclination angle $17\degree$. M87* shadow angular diameter $\theta_{sh}= 42\pm3\mu$as within $1 \sigma$ region, corresponding to the black dotted line $\theta_{sh}=39\mu$as, constrains the parameter space. The region above this line casts a shadow that is consistent with the M87* shadow size of EHT. The white region is forbidden for ($a/M,l/M$). The parameters along the red line describe an extremal black hole with degenerate horizons. The blue line represents a transition surface, with one positive and two degenerate horizons, whereas the  green line represents $\Delta(0)=0$. }\label{M871}
\end{center}
\end{figure}
\subsection{Observational constraints from the EHT results of M87*}
The EHT analysis suggested that, based on a priori known estimates for the mass and distance from stellar dynamics, the M87* shadow size is consistent within $17\%$ at a $68\%$ confidence interval of the size predicted from the Kerr black hole general relativistic-magneto-hydrodynamics (GRMHD) image \citep{EventHorizonTelescope:2019dse}. However, several other studies altogether have not entirely precluded the possibility of non-Kerr black holes \citep{Vincent:2020dij,Mizuno:2018lxz}. Using the M87* shadow angular size, constraints are placed on the second post-Newtonian metric coefficients, which were inaccessible in the earlier weak-field tests at the solar scale \citep{EventHorizonTelescope:2020qrl}. Therefore, it is both legitimate and timely to test the viability of the LQRBH using the M87* black hole shadow observations. We determine the constraints on the LMRBH parameters using the deviation from the circularity of the black hole shadow $\Delta C$ and the angular diameter $\theta_{sh}$, to validate the suitability of the LMRBH as a candidate for the M87* black hole.

Let the shadow boundary be described by the polar coordinates ($R(\varphi), \varphi$) and  the shadow has a center at ($X_c$, $Y_c$) with $X_c=(X_r-X_l)/2$ and $Y_c=0$. To assess the shadow deviation from a perfect circle, we define the dimensionless circularity deviation observable in terms of root-mean-square deviation from the average shadow radius as \citep{Johannsen:2010ru,Johannsen:2015qca, Kumar:2020yem}
\begin{equation}\label{circular}
\Delta C=\frac{1}{\bar{R}}\sqrt{\frac{1}{\pi}\int_0^{2\pi}\left(R(\varphi)-\bar{R}\right)^2d\varphi},
\end{equation}
where $\bar{R}$ is the shadow average radius defined as \citep{Johannsen:2010ru}
\begin{equation}
\bar{R}=\frac{1}{2\pi}\int_{0}^{2\pi} R(\varphi) d\varphi.
\end{equation}
with $\varphi\equiv \tan^{-1}[{Y}/({X-X_C})]$ being the subtended polar angle and $R(\varphi)=\sqrt{(X-X_{c})^2+(Y-Y_{c})^2}$ being the radial distance from the center ($X_c$, $Y_c$) of the shadow to any point ($X, Y$) on the boundary. The variables connected with black holes, such as their mass $M$, spin parameter $a$,  $l$, or inclination angle $\theta_0$, influence celestial coordinates, and as a result, also  the observables $\bar{R}$ and $\Delta C$. To determine the black hole parameter values that are most likely to be observed, we compare our theoretical prediction to the EHT observation and  use their result ($\Delta C\lesssim 0.1$) to impose constraints on the LMRBH parameter space while accounting for the inclination angle with respect to the observational line of sight to be $17$\textdegree \citep{Walker:2018muw}. 

Next, the shadow angular diameter  \citep{Kumar:2020owy,Kumar:2017tdw} is given by
\begin{equation}\label{angularDiameterEq}
\theta_{sh}=\frac{2}{d}\sqrt{\frac{A}{\pi}},
\end{equation}    
where $A$ is the black hole shadow area,as defined  in Equation~\ref{Area1}, and $d$ is the distance of M87* from the Earth. In Figure ~\ref{M871}, we demonstrate the LMRBH shadow angular diameter for $\theta_o=17$\textdegree~ as a function of $a$ and $l$. The M87* black hole shadow angular diameter within the $1\sigma$ bound, $39\mu as\leq\theta_{sh} \leq 45\mu as$, place a bound $0.0 \leq a\leq 0.52757 M$, whereas all values of $l$ are allowed. For $a > 0.52757 M$, there is an  upper limit on the value of $l$ which depends on the value of $a$ (see Figure  \ref{M871}). Thus, in this constrained parameter space, the M87* can be LMRBH spacetime. Since there is a number of possible parameter points ($a$, $l$) within the confined parameter space, the LMRBH's compatibility with the M87* observations demonstrates that they can be excellent candidates for astrophysical black holes. Circularity deviation is shown in Figure ~\ref{deltaC} for $\theta_o=17$\textdegree. The M87* bound $\Delta C\leq 0.10$ is satisfied by the LMRBH for the entire parameter space. This is because the rotating black hole shadows are nearly circular for small inclination angle. 
\begin{figure}
\begin{center}
	\includegraphics[scale=0.75]{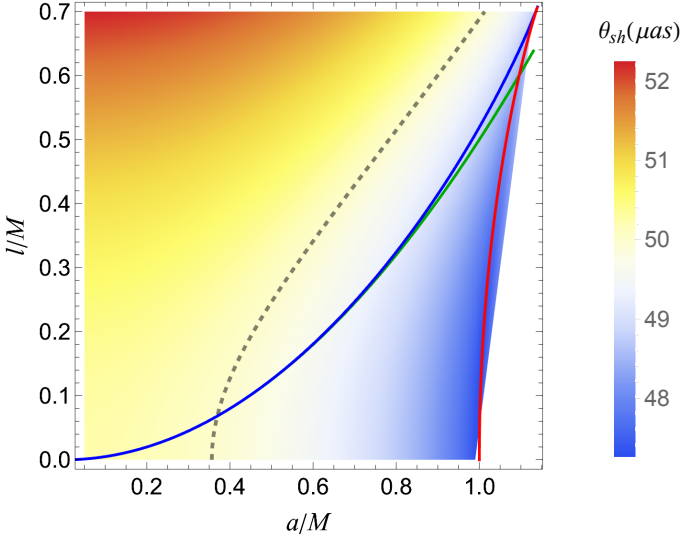}
	\caption{LMRBH shadow angular diameter $\theta_{sh}$, in units of $\mu$as, as a function of ($a, l$) for an inclination angle $50\degree$. Black-dashed line corresponds to   $\theta_{sh}=50\mu$as. The region below this line casts a shadow that is consistent with the Sgr A* shadow size of EHT. The white region is forbidden for ($a/M,l/M$). The parameters along the red line describe an extremal black hole with degenerate horizons. The blue line is a transition surface, with one positive and two degenerate horizons, whereas the  green line represents $\Delta(0)=0$. }\label{sgr1}
\end{center}	
\end{figure}
\begin{figure}
	\includegraphics[scale=0.75]{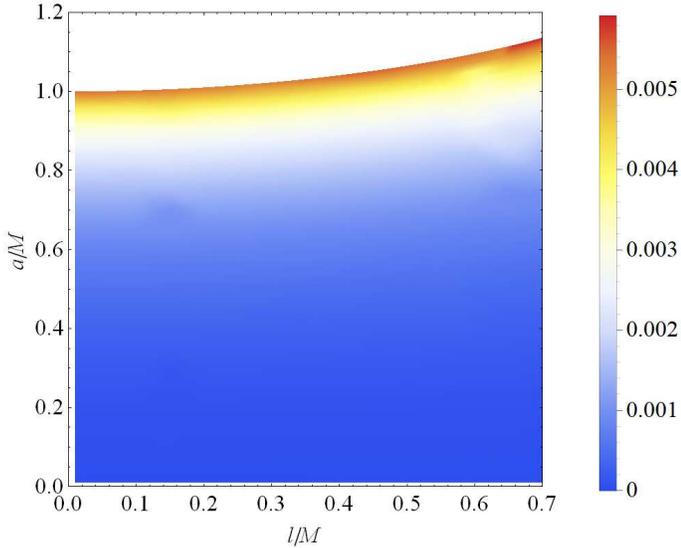}
\caption{Circularity deviation $\Delta C$ as a function of $(a, l)$.  }\label{deltaC}
\end{figure}

\subsection{Observational constraints from the EHT results of Sgr A*}
Based on the 2017 VLBI observing campaign at 1.3 mm wavelength, the EHT collaboration \citep{EventHorizonTelescope:2022exc,EventHorizonTelescope:2022urf,EventHorizonTelescope:2022vjs,EventHorizonTelescope:2022wok,EventHorizonTelescope:2022xnr, EventHorizonTelescope:2022xqj}  published the shadow data for the Sgr A* black hole. Sgr~A* black hole shadow images have advantages in testing the nature of an astrophysical black hole (i) Sgr~A* probes a $10^{6}$ order of higher curvature than the M87* and (ii) independent prior estimates for mass-to-distance ratio are used for Sgr~A*. The shadow images were created using a variety of imaging and modeling techniques, and they are astonishingly similar in features.
\begin{figure}
\begin{center}
	\includegraphics[scale=0.75]{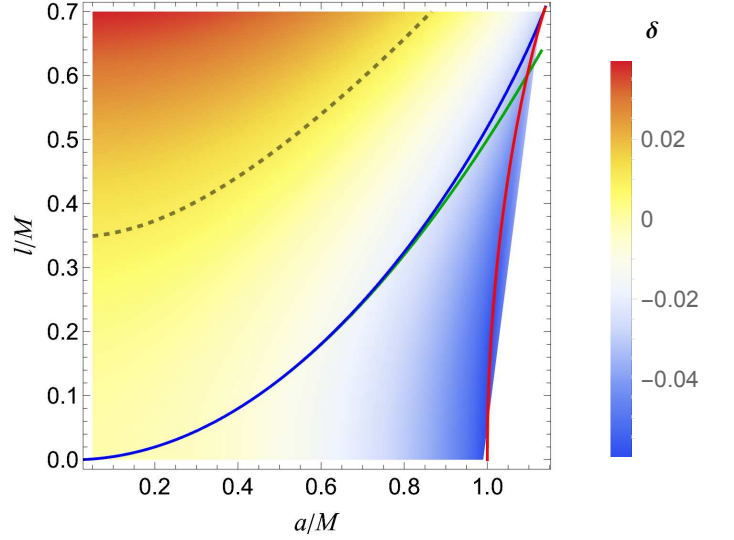}
	\caption{ LMRBH shadow angular diameter deviation from that of a Schwarzschild black hole as a function of ($a,l$). The Sgr A* black hole shadow restrictions by EHT are indicated by the black-dashed line at $\delta = 0.01$. The white region is forbidden for ($a/M,l/M$). The parameters along the red line describe an extremal black hole with degenerate horizons. The blue line is a transition surface, with one positive and two degenerate horizons, whereas the  green line represents $\Delta(0)=0$.}\label{sgr2}
\end{center}	
\end{figure}
We will use the EHT bounds on the two observables, shadow angular diameter $\theta_{sh}$ and Schwarzschild shadow deviation  $\delta$, to set constraints on the LMRBH parameters. We determine $\theta_{sh}$, which, in addition to other black hole parameters, relies on the mass $M$, the distance $d$ of the black hole, LQG parameter $l$, and inclination $\theta_0$ as depicted in Figure~\ref{sgr1}. We show that $41.7\mu as\leq \theta_{sh} \leq 55.7\mu as$, i.e., within the $1 \sigma$ region of the SgrA* shadow angular diameter,  is satisfied for entire parameter space in the case of the LMRBH. Moreover, EHT used three independent imaging algorithms, namely, \texttt{ eht-imaging}, \texttt{SIMLI}, \texttt{DIFMAP}, to determine the Sgr~A* shadow's average diameter $46.9\mu $as $\leq \theta_{sh} \leq 50~\mu$as. 
This range of  angular diameter strongly constrains the parameters  $0.356355 M \leq a \leq a_{c}$ and  $0 \leq l \leq l_u$ at $\theta_o=50$\textdegree. $a_{c}$ are the critical values of parameter $a$ represented  by the red line in Figure ~\ref{sgr1} and $l_u$ is the maximum value of $l$ which depends on $a$. For this constrained parameter range, the LMRBH shadows are consistent with the shadow of Sgr A*.

%Moreover, for a black hole of a specific angular size, the angular diameter of the shadow can be used to gauge the characteristics of the black hole metric and assess how well it agrees with the Kerr solution of general relativity  \cite{EventHorizonTelescope:2022xnr,EventHorizonTelescope:2022xqj}.  Infact 
The Schwarzschild shadow deviation ($\delta$),  assesses the  disparity between the shadow angular diameter of the LMRBH, $\theta_{sh}$, and the Schwarzschild shadow diameter,  $\theta_{sh, Sch}$, and is given as \citep{EventHorizonTelescope:2022xnr,EventHorizonTelescope:2022xqj}
\begin{equation}
\delta=\frac{\theta_{sh}}{\theta_{sh, Sch}}-1.
\end{equation}
For Kerr black hole with $a\leq M$, the Schwarzschild shadow deviation lies at $-0.075\leq  \delta \leq 0$ as the inclination varies from 0 to $\pi/2$.
EHT used the two separate priors for the Sgr~A* angular size from the Keck and Very Large Telescope Interferometer (VLTI) observations to estimate the bounds on the fraction deviation observable $\delta$  \citep{EventHorizonTelescope:2022xnr,EventHorizonTelescope:2022urf}
\begin{align}
\delta_{Sgr}= \begin{dcases*} -0.08^{+0.09}_{-0.09}\;\;\;\;\; & \text{VLTI}\\
-0.04^{+0.09}_{-0.10}\;\;\;\;\; &\text{Keck}	 
\end{dcases*} 
\end{align}
Thus, modeling Sgr A* as an LMRBH, the entire parameter space in the case of the LMRBH is satisfied for inferred bound Keck (-0.14,0.05). On the other hand, the VLTI bound (-0.17,0.01), constrains the parameters ($a,l$) such that for $0< l\leq 0.347851M\; (l\leq 2\times 10^6$ Km), the allowed range of $a$ is $(0,1.0307M)$. If $a>1.0307M$, there is an upper bound on the value of $l$ which is not specific and depends on the value of '$a$', for e.g., at $a=0.5094M$, $l \leq 0.4864M$ and at $a=0.8399 M$, $l \leq 0.6758 M$ ( Figure~\ref{sgr2}).

\section{Conclusion}\label{Sec-6}
The scarcity of rotating black hole models in the LQG  restricts the progress of testing LQG from observations, like the EHT observation of M87* and SgrA*. Furthermore, because of the rapidly growing astronomical observations of rotating black
holes, starting from a nonrotating LQG seed metric (Equation (\ref{metric2})) and using a modified NJA generating technique we obtained a rotating LMRBH in an earlier work \citep{Kumar:2022vfg}. The LMRBH metric possesses a Kerr-like form and has several exciting properties \citep{Kumar:2022vfg}. It is nonsingular everywhere and reduces to the Kerr solution when the quantum effects are switched off ($l=0$). Depending on values of parameters $(M,\;a,\;l)$, the LMRBH can describe an NH, a generic regular black hole with Cauchy and event horizons (BH-II), or a black hole with one or three horizons (BH-I or BH-III). 
%We show that EHT observations have entirely ruled out the possibility of LMRBH being an NH without a horizon.

Here, we are interested in the shadow and analyzed the LMRBH  to find that the Hamilton-Jacobi equations are separable, resulting in null geodesic equations in the first-order differential form.   In Figure  \ref{fig3}, we depict the shadow cast by the LMRBH in the parameter space to find black hole shadow size  increases monotonically, and the shape becomes more distorted with an increasing $l$.  For the given shadow observables, we have estimated parameters associated with the LMRBH by two popular methods, assuming that the observer is in the equatorial plane. We show that by using these shadow observables one can ultimately determine information about an LMRBH.

The supermassive black holes,  M87* and Sgr A*, at the center of the Milky Way and M87, are a superior and realistic laboratory for testing the strong-field predictions of GR. 
The EHT collaboration has released the first horizon-scale images of both M87* and Sgr A*, and the EHT results are consistent with the prediction on the Kerr metric, and there is no evidence for any breach of the theory of GR. Here, we used the EHT observation of the black hole shadow in  M87* and Sgr A* to place constraints on the deviation parameters associated with LMRBH.  

To place constraints on the deviation parameter $l$,  we employ the angular size and asymmetry results of the EHT for the M87* black hole and the EHT bounds on the Sgr A* shadow angular diameter and Schwarzschild shadow deviation from the Sgr A* results. The allowed range of M87* angular shadow diameter within $1\sigma$ region, i.e., $39\mu$as$\leq \theta_{sh}\leq 45\mu as$ is possible for the entire range of $l$ and $0.0 \leq a\leq 0.52757 M$. However, if $a > 0.52757 M$, then there is an upper limit on $l$ which depends on  $a$. Likewise, we show that the Sgr A* shadow angular diameter within $1\sigma$ credible region, $41.7\mu as\leq \theta_{sh} \leq 55.7\mu as$, is satisfied for the entire parameter space in the case of LMRBH. But, the EHT observation also employed three different techniques to determine that the shadow's average measured angular diameter is within the range $\theta_{sh} \in (46.9, 50)~\mu$as which strongly constrains the parameters, such that $0.356355 M \leq a \leq a_{c}$ and  $0<l<l_u$  at $\theta_o=50$\textdegree  is allowed. Here, $a_{c}$ and $l_u$ are, respectively,  the critical values of parameter $a$ represented  by the red line in Figure ~\ref{sgr1} and maximum value of $l$. Further, modeling Sgr A* as an LMRBH, the entire parameter space with the LMRBH is obeyed for the inferred  Keck bound (-0.14,0.05). The VLTI bound (-0.17,0.01), constrains the parameters ($a,l$) such that for $0 \leq l \leq 0.347851M$, the allowed range of $a$ is $(0,1.0307M)$. For $a>1.0307M$, the maximum value of $l$  depends on '$a$'. The shadow circularity deviation $\Delta C$ bound for M87* black hole allows the entire parameter space for LMRBH due to the low inclination angle, whereas the $\Delta C$ bound for Sgr A* is not available.

The main restriction of our proposition is that the LMRBH metric does not result from a direct loop quantization of the Kerr spacetime.  However, we note that the existence of the region BH-I  has similar features of nonrotating LQG black holes (\ref{metric1}) (see the Penrose Diagram in Figure  \ref{pen4}) and the LMRBH also provides the singularity resolution of Kerr black holes. Thus, it is pragmatic to expect that the LMRBH can capture some aspects of the  effective regular spacetime description of LQG. 
 
\section{Acknowledgments}  S.U.I and S.G.G. are supported by SERB-DST through project No.~CRG/2021/005771. J.K. would like to thank CSIR for providing SRF. R.K.W. would like to thank  the University of KwaZulu-Natal and the NRF for the postdoctoral research fellowship.
\bibliography{LQG}
\bibliographystyle{aasjournal}
\end{document}